\documentclass[11pt, a4paper]{article}
\usepackage[text={8in,11in},margin=1.0in,includefoot]{geometry}
\usepackage{secdot}
\sectiondot{subsection}
\usepackage{doc}
\usepackage{booktabs,caption}
\usepackage{array}
\newcolumntype{P}[1]{>{\centering\arraybackslash}p{#1}}
\newcolumntype{M}[1]{>{\centering\arraybackslash}m{#1}}

\usepackage{url}
\usepackage{graphicx}
\usepackage{epstopdf}
\usepackage{tabularx}
\usepackage{amsmath}
\usepackage{pdflscape}
\usepackage{float}
\usepackage{placeins}
\usepackage{caption}
\usepackage{booktabs}
\usepackage{cite}
\usepackage{multirow}
\usepackage{longtable}
\usepackage[export]{adjustbox}
\usepackage[english]{babel}
\usepackage[autostyle]{csquotes}
\usepackage{array,longtable}
\renewcommand*{\arraystretch}{1.5}

\restylefloat{table}
\numberwithin{equation}{section}
\title{Properties of $J^{P}=1/2^{+}$ baryon octets at low energy}
\author{$^1*$Amanpreet Kaur, Pallavi Gupta, Alka Upadhyay
\\\small{\it $^1$ School of Physics and Material Science},\\\small{\it Thapar University,
Patiala, Punjab-147004}\\\small{E-mail:
$^1*$amanpreet.kaur9074@yahoo.com}}

\begin{document}
\date{}

\maketitle
\begin{abstract} The statistical model in combination with detailed balance principle
is able to phenomenologically calculate and analyze spin and flavor
dependent properties like magnetic moments (with effective masses,
effective charge, with both effective mass and effective charge),
quark spin polarization and distribution, strangeness suppression
factor and $\overline{d}-\overline{u}$ asymmetry incorporating
strange sea. The $s\overline{s}$ in the sea is said to be generated
via the basic quark mechanism but suppressed by the strange quark
mass factor $m_{s}>m_{u,d}$. The magnetic moments of the octet
baryons are analyzed within the statistical model, by putting
emphasis on the SU(3) symmetry breaking effects generated by the
mass difference between the strange and non strange quarks. The work
presented here assume hadrons with a sea having admixture of
quark-gluon Fock states. The results obtained have been compared
with theoretical models and experimental data.

\vspace{0.2cm}
\end{abstract}
\section{Introduction}
Last two decades has witnessed the phenomenal discoveries in
particle physics that were celebrated and honored by two Nobel
Prizes. Francois Englert and Peter Higgs were jointly awarded with
Nobel Prize in 2013 \enquote{for the theoretical discovery of a
mechanism that contributes to our understanding of the origin of
mass of subatomic particles, and which recently was confirmed
through the discovery of the predicted fundamental particle, by the
ATLAS and CMS experiments at CERN's Large Hadron Collider}. The
Nobel Prize of 2015 was awarded to Arthur McDonald and Takaaki
Kajita \enquote{for the discovery of neutrino oscillations, which
shows that neutrinos have mass}, contradicting the Standard Model
which states neutrinos are massless particles. Thus, physics beyond
standard model can be studied by working on these new achievements.
Also, a new state of matter called \enquote{pentaquarks} was
reported in LHCb experiment at LHC bringing a glorious triumph in
the study of baryon spectroscopy and \enquote{studying its
properties may allow us to understand better how ordinary matter,
the protons and neutrons from which we are all made, is
constituted}.

Brodsky, Hoyer, Peterson and Sakai (BHPS) model\cite{1} suggested
that there are two types of quark contribution to nucleonic sea:
intrinsic and extrinsic. Intrinsic sea (valencelike) originates
through nonperturbative fluctuations of nucleonic state to five
quark state ($uudq\overline{q}$) and extrinsic sea (sealike) is
produced perturbatively in the process of gluons splitting into
$q\overline{q}$ pairs \cite{2,3}. There is wide range of results
available for intrinsic light quark distributions from meson cloud
model \cite{4,4a,4b,4c}, chiral quark model \cite{5,5a} etc. Parton
distribution functions (PDFs) has also been successful in extracting
the probabilities of intrinsic charm quark in the nucleon
\cite{6,6a}.

The study of internal structure of hadrons is still one of the
unresolved issue despite being studied extensively over the globe in
the form of experiments and theoretical developments over the past
50 years. The study of hadrons is a complication for physicists due
to relativistic and non-perturbative nature of its constituents. The
presence of quark-gluon interaction in valence quarks implies that
$q\overline{q}$ pairs can be produced perturbatively by gluons
emitted from valence quarks itself which are usually termed as
"sea"\cite{0}. The sea could be consisting of light-quarks as well
as heavy ones and plays an important role in visualizing the
hadronic structure. The distribution of $\overline{u}$ and
$\overline{d}$ in sea have large asymmetry according to observations
in deep inelastic scattering\cite{7,7a} and Drell-Yan experiments
\cite{8,8a}. The hadrons made of valence quarks and gluonic degrees
of freedom in terms of "sea" have been modeled diversely to
interpret its observed properties like magnetic moments, spin
distributions, quadrupole moments etc. This class of models assumes
nucleon consist of valence quarks surrounded by a "sea" which, in
general, contains gluons and virtual quark-antiquark pairs, and is
characterized by its total quantum number \cite{9}. The domains of
validity and stability of the results obtained can be checked by
calculating maximum number of hadronic parameters or properties
within these models. Various nucleonic parameters have been
calculated theoretically using chiral quark soliton model \cite{9a},
Skyrme model \cite{9b}, sum rules etc \cite{9c}.

An attempt by Zhang et. al. \cite{10} was made to understand the
flavor asymmetry of the proton in detailed balance model. The idea
was, that proton is taken as ensemble of quark gluon Fock states and
any two nearby Fock states should be balanced with each other under
the principle of detailed balance in a way \cite{10a,10b}:
\begin{gather*}
\rho_{A}R_{A\rightarrow B}=\rho_{B}R_{B\rightarrow A}
\end{gather*} where $\rho_{A}$ and $\rho_{B}$ are the probabilities
of finding the proton in state of A and B respectively,
$R_{A\rightarrow B}$ and $R_{B\rightarrow A}$ are the transition
probabilities of A to B and B to A respectively. The probabilities
of finding every Fock state inside the proton were obtained in the
same process. The detailed balance model was used to look into
statistical effects of the nucleon and $\overline{d}-\overline{u}$
asymmetry. The model generated $\overline{d}-\overline{u}$= 0.124
which was in agreement with the predictions of E866/NuSea result of
$0.118\pm0.012$. We have computed $\overline{d}-\overline{u}$
asymmetry for all other octet baryons using the relation \cite{14}:
\begin{gather}
\overline{d}-\overline{u}=
([\sum\limits_{\substack{j=0,k=0\\i=l=0}}^{j=2,k=3}
\rho_{i,j,l,k}]-[\sum\limits_{\substack{i=0,k=0\\j=l=0}}^{i=2,k=3}
\rho_{i,j,l,k}])
\end{gather}The values for flavor asymmetry are shown
in table 1 followed by list of average number of partons for all
$J^{P}=\frac{1}{2}^{+}$ octet members. The total number of intrinsic
partons inside every baryon of octet can be calculated as:
$N^{'}=u_{val}+d_{val}+s_{val}+\overline{u}u_{sea}+\overline{d}d_{sea}+\overline{s}s_{sea}+g$
and from the normalization condition,
$\sum_{i,j,l,k}\rho_{i,j,l,k}=1$, we can find the average number of
partons for all baryons.


\setlength{\tabcolsep}{0.9em} %
{\renewcommand{\arraystretch}{1.9}%
\FloatBarrier
\begin{table}[H]
\caption{$\overline{d}-\overline{u}$ asymmetry and average number of
partons for octet baryons.}
\begin{center}
\begin{tabular}{|cccccc|}
  \hline
  $J^{P}=\frac{1}{2}^{+}$ octets &$\overline{u}$&$\overline{d}$& $\overline{d}-\overline{u}^{*}$ & Data$^{*}$& Avg. no. of partons \\\hline
  \hline
p &0.33&0.46 &0.134  &  0.124 \cite{10c}& 4.84\\
n &0.46&0.33& -0.134 & -& 4.84  \\
$\Lambda^{0}$ &0.44 & 0.44&0& 0 &4.73\\
$\Sigma^{+}$&0.31&0.71&0.39 &  0.410 \cite{15}&4.87   \\
$\Sigma^{0}$&0.44& 0.44&0& 0 &4.74 \\
$\Sigma^{-}$&0.71  & 0.31&-0.39& -&4.87 \\
$\Xi^{0}$ & 0.43 & 0.70&0.26 &0.27 \cite{15}&4.85\\
$\Xi^{-}$ & 0.70  & 0.43&-0.26&-&4.85 \\
\hline
\end{tabular}
\end{center}
\end{table}
\FloatBarrier Later, J. P Singh et. al. \cite{9} constructed Fock
states of nucleon having specific color and spin quantum numbers
with definite symmetry properties, using statistical ideas. With
this approximation, they studied the quarks contribution to the spin
of the nucleons, the ratio of the magnetic moments of the nucleons,
their weak decay constant and the ratio of SU(3) reduced matrix
elements for the axial current. We in this paper, have used the
concept for other octet particles to reproduce the flavor asymmetry,
quark spin polarization and distribution, magnetic moments (with
various modifications) and SU(3) symmetry breaking in magnetic
moments for octet particles, which may help in unraveling the
hadronic structure. We have also computed the strangeness
suppression for octet particles in the framework of the statistical
model. Because masses of u and d quarks are fairly small compared
with a typical energy scale in the deep inelastic scattering, the
splitting processes are expected to occur almost equally for these
quarks but a constraint is needed to be applied for strange
quark-antiquark pair. The contribution of the $s\overline{s}$ pair
is taken into account by applying a constraint resulting from gluon
free energy distribution \cite{10b}. Orbital motion of the three
valence quarks have not been taken into account, however, some of
the previous studies shows inclusion of orbital motion preserves the
success in describing the magnetic moments\cite{11} , spin-averaged
structure function and violation of Gottfried sum rules\cite{12}.
The motivation of this paper is to generalize the conjecture of
detailed balance principle and statistical model, used earlier in
the calculation of some low energy properties of octet particles.

\section{Magnetic moments using constituent quark masses and Spin distribution of hyperons}
Out of many approaches and models such as the lattice QCD,
quark-diquark model, chiral constituent model, potential model, QCD
sum rules etc., available for the three body systems, we employ here
the statistical approach to study the bayons in
$J^{P}=\frac{1}{2}^{+}$ state. Magnetic moments is low energy and
long distance phenomenon. The magnetic moments of
$J^{P}=\frac{1}{2}^{+}$ baryons are computed using the spin-flavor
wave functions of the constituting quarks. Here, we have
incorporated the effect of (a) quark effective masses, (b) quark
effective charges and (c) both i.e quark effective mass plus
effective charge, to compute the magnetic moments.
The quark magnetic moments in terms of effective quark masses can be
written as \cite{12a}:
\begin{equation}
\begin{split}
\mu_{u}^{eff}=2[1-(\Delta M/M_{B})]\mu_{N},\\
\mu_{d}^{eff}=-[1-(\Delta M/M_{B})]\mu_{N},\\
\mu_{s}^{eff}=-M_{u}/M_{s}[1-(\Delta M/M_{B})]\mu_{N}
\end{split}
\end{equation}
where $M_{B}$ is the mass of the baryon obtained additively from the
quark masses and $\Delta M$ is the mass difference between the
experimental value and $M_{B}$. The expressions for magnetic moments
for all octet particles can be expressed in terms of effective quark
magnetic moments ($\mu_{u}^{eff},\mu_{d}^{eff},\mu_{s}^{eff}$) and
two parameters $\alpha$ and $\beta$ statistically, as shown in table
3. The values of $\alpha$ and $\beta$ are calculated by including
the strange quark-antiquark pairs in sea \cite{121}. For baryons at
ground state, the magnetic moments is a vector sum of quark magnetic
moments,
\begin{equation}
\label{eq:wavefunction18}
\mu_{baryon}=\sum_{i=1,2,3}\mu_{i}\sigma_{i}
\end{equation}
where $\sigma_{i}$ is the pauli matrix representing the spin term of
$i^{th}$ quark and $\mu_{i}$ represents magnitude of quark magnetic
moments and therefore, values of magnetic moments are different for
different baryons. Also,
\begin{equation}
\label{eq:wavefunction19} \mu_{i}=\frac{e_{i}}{2m_{i}}
\end{equation}
for i = u, d, s and $e_{i}$ represents the quark charge. We have
applied the magnetic moment operator
$(\widehat{O}=\mu_{i}\sigma_{i})$ to the octet wave function in a
way:
\begin{equation}
\label{eq:wavefunction20}
\mu_{B}=\sum\limits_{\substack{u,d,s}}\langle
\Psi_{B}|\frac{e_{i}\sigma_{Z}^{i}}{2m_{eff}}|\Psi_{B} \rangle
\end{equation}
In general,
\begin{equation}
\begin{split}
\langle\Phi^{(\uparrow)}_{1/2}|\hat{O}|\Phi^{(\uparrow)}_{1/2}\rangle=\frac{1}{N^{2}}[\langle\Phi^{(1/2\uparrow)}_{1}|\hat{O}|\Phi^{(1/2\uparrow)}_{1}\rangle+
\sum\limits_{\substack{i=8,10}}
a_{i}^{2}\langle\Phi^{(1/2\uparrow)}_{i}|\hat{O}|\Phi^{(1/2\uparrow)}_{i}\rangle+\sum\limits_{\substack{i=1,8,10}}
b_{i}^{2}\langle\Phi^{(1/2\uparrow)}_{bi}|\hat{O}|\Phi^{(1/2\uparrow)}_{bi}\rangle+\\2b_{8}c_{8}\langle\Phi^{(1/2\uparrow)}_{b8}|\hat{O}|\Phi^{(1/2\uparrow)}_{c8}\rangle+
c_{8}^{2}\langle\Phi^{(1/2\uparrow)}_{c8}|\hat{O}|\Phi^{(1/2\uparrow)}_{c8}\rangle+d_{8}^{2}\langle\Phi^{(1/2\uparrow)}_{d8}|\hat{O}|\Phi^{(1/2\uparrow)}_{d8}\rangle]
\end{split}
\end{equation}
The magnetic moment relations obtained after applying operator for
$J^{P}=\frac{1}{2}^{+}$ particles in terms of parameters $\alpha$,
$\beta$ and quark effective masses are shown in column 1 of table 2.
We repeat our computations by varying the effective masses of quarks
(in MeV) from 370 to 390 for u and d quarks and 500 to 530 for
strange quark, to have an idea about the most suitable set of
effective quark masses that yields the magnetic moments of baryons.
As the values of effective masses are model dependant so the
magnetic moments of quarks are also model dependent and one has to
take their values consistent with the constituent quark masses.

In addition to the calculation of magnetic moments with effective
mass, we have also computed magnetic moments with effective charge.
Here, we have applied the magnetic moment operator
$(\widehat{O}=\mu_{i}\sigma_{i})$ to the octet wave function in a
way:
\begin{equation}
\label{eq:wavefunction20}
\mu_{B}=\sum\limits_{\substack{u,d,s}}\langle
\Psi_{B}|\frac{e_{i}^{(eff)}\sigma_{Z}^{i}}{2m_{eff}}|\Psi_{B}
\rangle
\end{equation}
We have taken the effective charge to depend linearly on the charge
of the shielding quarks. Thus, the effective charge $e_{a}$ of quark
a in the baryon B(a,b,c) is written as \cite{12b}:
\begin{equation}
e_{a}^{B}=e_{a}+\alpha_{ab}e_{b}+\alpha_{ac}e_{c}
\end{equation} where $e_{a}$ is the bare charge of quark a. Taking
isospin symmetry, we have
\begin{gather*}
\alpha_{uu}=\alpha_{ud}=\alpha_{dd}=\beta\\
\alpha_{us}=\alpha_{ds}=\alpha\\
\alpha_{ss}=\gamma
\end{gather*}The screened quark charges for various baryons in terms of the three
parameters $\alpha, \beta, \gamma$ can be expressed as:
\begin{equation}
\begin{split}
e_{u}^{p}=\frac{2}{3}(1+\frac{1}{2}\beta),
e_{d}^{p}=-\frac{1}{3}(1-4\beta),\\ e_{u}^{n}=\frac{2}{3}(1-\beta),
e_{d}^{n}=-\frac{1}{3}(1-\beta)...
\end{split}
\end{equation}
Substituting these values of effective charge and applying the
magnetic moment operator to wave function as in equation (2.4) and
using the unknown parameters in effective charge relations i.e.
$\alpha$= 0.248, $\beta$=0.025, $\gamma$= 0.018 to 0.029, as input,
we determine the magnetic moments and are shown in table 3.

SU(3) symmetry breaking effects is also applied in sea and valence
quarks. This breaking is due to the mass difference between strange
and non-strange light quarks. Symmetry breaking is applied to the
magnetic moments in the form of the parameter $r=m/m_{s}$ \cite{16}
where m is the mass of u and d quarks. The value of \enquote{r} lies
in the range 0.70 to 0.78, depending upon the respective effective
quark masses. The results for magnetic moments with SU(3) broken
symmetry for hyperons are shown in table 4.

Recently, CLAS Collaboration \cite{12c} reported the strangeness
suppression in the proton from the production rates of baryon-meson
states in exclusive reactions, i.e. without the production of an
intermediate baryon resonance. We are also interested in measuring
how often, compared with pairs of light quarks, strange quarks are
made. For this purpose, we define the strangeness suppression factor
as
$\lambda_{s}=\frac{2(s\overline{s})}{(u\overline{u}+d\overline{d})}$.
This ratio implies the existence of strange quarks in sea. So, we
have calculated the strangeness suppression factor for all particles
in $J^{P}=\frac{1}{2}^{+}$ state in the framework of principle of
detail balance. The calculation of strangeness suppression factor
includes various sub processes like $g\Leftrightarrow
q\overline{q}$, $g\Leftrightarrow gg$, $q\Leftrightarrow qg$. The
results are presented in Table 5. The value for the strangeness
suppression factor is in good agreement with both the values
determined in exclusive reactions and in high-energy production.

\setlength{\tabcolsep}{0.6em} %
{\renewcommand{\arraystretch}{1.8}%
\FloatBarrier
\begin{table}[H]
\tiny
 \caption{Magnetic moments of $J^{P}=\frac{1}{2}^{+}$ baryons
with (a) effective quark masses, (b) quark effective charge and (c)
both i.e quark effective mass plus quark effective mass. The results
of magnetic moments with various modifications are shown taking two
and three gluons in sea respectively.}
\begin{center}
\begin{tabular}{|p{3.6cm}|c|c|c|c|c|c|c|}
\hline
&\multicolumn{6}{c|}{Magnetic moments $(\mu_{N})$}&\\
 \cline{2-7}
Baryon octet magnetic moments &\multicolumn{2}{p{1.1cm}|}%
 {\centering With effective quark mass}&\multicolumn{2}{p{1.8cm}|}%
 {With effective quark charge }&\multicolumn{2}{p{1.9cm}|}%
{ \centering With effective quark mass+ effective quark charge}&\\
  \cline{2-7}
&(2 gluons)&(3 gluons)&(2 gluons)&(3 gluons)&(2 gluons)&(3 gluons)
&\multicolumn{1}{p{1.1cm}|}%
{Exp. Results \cite{16a}}
\\
\hline \hline


$\mu_{p}=3(\mu_{u}^{eff}\alpha-\mu_{d}^{eff}\beta)$&2.79&2.29&1.81&1.73&2.74&2.62&2.79\\
 $\mu_{n}=3(\mu_{d}^{eff}\alpha-\mu_{u}^{eff}\beta)$&-1.83&-1.50&-0.85&-0.84&-1.38&-1.37&-1.91\\

  $\mu_{\Lambda}=\frac{1}{2}(\alpha-4\beta)(\mu_{u}^{eff}+\mu_{d}^{eff})+(2\alpha+\beta)\mu_{s}^{eff}$&-0.634&-0.60&-0.41&-0.36&-0.59&-0.52&-0.613\\
  $\mu_{\Sigma^{+}}=3(\mu_{u}^{eff}\alpha-\mu_{s}^{eff}\beta)$&2.464&2.11&1.99&1.78&3.0&2.77&2.458\\
   $\mu_{\Sigma^{0}}= \frac{3}{2}(\mu_{u}^{eff}\alpha+\mu_{d}^{eff}\alpha-2\mu_{s}^{eff}\beta)$&0.775&0.680&0.82&0.76&1.31&1.19&0.775\\
  $\mu_{\Sigma^{-}}=3(\mu_{d}^{eff}\alpha-\mu_{s}^{eff}\beta)$&-0.974&-0.82&-0.83&-0.75&-1.29&-1.17&-1.160\\
   $\mu_{\Xi^{0}}=3(\mu_{s}^{eff}\alpha-\mu_{u}^{eff}\beta)$&-1.388&-1.203&-0.933&-0.84&-1.36&-1.23&-1.250\\
    $\mu_{\Xi^{-}}=3(\mu_{s}^{eff}\alpha-\mu_{d}^{eff}\beta)$ &-0.615&-0.53&-0.422&-0.403&-0.57&-0.55&-0.6507\\
 \hline
\hline
\end{tabular}
\end{center}
\end{table}
\FloatBarrier

\setlength{\tabcolsep}{0.9em} %
{\renewcommand{\arraystretch}{1.9}%
\FloatBarrier
\begin{table}[H]
\caption{Magnetic moments of particles with SU(3) symmetry breaking,
SU(3) symmetry and comparison with experimental data available.}
\begin{center}
\begin{tabular}{|c|c|c|c|}
  \hline
  &\multicolumn{3}{c|}{Magnetic moments $(\mu_{N})$}\\
 \cline{2-4}
{Baryon octet magnetic moments} &\multicolumn{1}{p{2.1cm}|}%
 {\centering With SU(3) symmetry breaking}&\multicolumn{1}
 {p{2.1cm}|}%
 {With SU(3) symmetry}&\multicolumn{1}
 {p{1.8cm}|}%
 {Exp. Results \cite{16a}}
\\\hline
$\Lambda^{0}$ &-0.44 & -0.634&-0.613\\
$\Sigma^{+}$& 2.40&2.464&2.458  \\
$\Sigma^{0}$&0.644&0.775&0.775  \\
$\Sigma^{-}$&-1.018& -0.974&-1.160 \\
$\Xi^{0}$ & -1.063& -1.388&-1.250\\
$\Xi^{-}$ & -0.35 &-0.615&-0.6507 \\
\hline
\end{tabular}
\end{center}
\end{table}
\FloatBarrier

\setlength{\tabcolsep}{0.9em} %
{\renewcommand{\arraystretch}{1.9}%
\FloatBarrier
\begin{table}[H]
\caption{Strangeness suppression for all particles in octet are
shown in this table. The values in the second row with single *
shows the results from UQM \cite{12d} and with double ** are the
experimental results \cite{12c} for strangeness suppression in
proton. }
\begin{center}
\begin{tabular}{cccc}
  \hline
  Baryon octet&$\frac{s\overline{s}}{d\overline{d}}$&$\frac{u\overline{u}}{d\overline{d}}$& $\lambda_{s}=\frac{2(s\overline{s})}{(u\overline{u}+d\overline{d})}$\\\hline
  \hline
p&0.32/$0.26^{*}$/$0.22^{**}$&0.71/$0.57^{*}$/$0.74^{**}$&0.38/$0.34^{*}$/$0.29^{**}$\\
n&0.46&1.40&0.38\\
$\Lambda^{0}$ &0.28&1 &0.29\\
$\Sigma^{+}$&0.19 &0.43&0.277  \\
$\Sigma^{0}$&0.28&1&0.28  \\
$\Sigma^{-}$&0.45&2.26 &0.276 \\
$\Xi^{0}$ &0.17&0.62 &0.21\\
$\Xi^{-}$ & 0.28 &1.61&0.21 \\
\hline
\end{tabular}
\end{center}
\end{table}
\FloatBarrier

The individual spin polarization due to quarks for hyperons are
calculated and compared with data of other available model. In
application, the individual spin polarization is defined as, $\Delta
q=n(q\uparrow)-n(q\downarrow)+n(\overline{q}\uparrow)-n(\overline{q}\downarrow)$
for q=u, d, s, where $n(q\uparrow)$ is the number of spin-up and
$n(q\downarrow)$ is the number of spin down quarks of flavor q for
both quarks and anti-quarks. In addition, total spin distributions
of baryon is also determined by applying the operator
$\frac{1}{2}e_{i}^{2}\sigma_{Z}^{i}$ where $e_{i}$ and
$\sigma_{Z}^{i}$ are the charge of quark and spin projection
operator respectively, to the baryonic wave function. The results
from statistical model and comparison with other models are shown in
table 2.

A well known and important problem for physicists over last 20 years
i.e. proton spin crisis suggested that proton's spin is built from
constituent quarks plus sea of quark-antiquark pairs and gluons.
Deep inelastic experiments and European Muon Collaboration predicted
that total spin of proton is very little contributed by quark's
intrinsic spin, which was contrary to the results of
non-relativistic quark model. The whole story led the
phenomenologist as well as experimentalist to think beyond the
already known facts. Various experiments \cite{17} measured spin
structure function $(g_{1})$ at x=0.1 to x=0.01 concluded that, the
proton is a system of three massive constituent quarks interacting
self-consistently with cloud of virtual pions and condensates
generated from spontaneous breaking of chiral symmetry between left
and right handed quarks. On the other other hand, when probed at
high resolution, the structure of proton seems to be combination of
three valence quarks plus sea of quark-antiquark pairs and gluons.
Thus, we conclude that nucleonic spin is distributed among gluons,
valence and sea quarks plus their angular momenta.

\setlength{\tabcolsep}{0.2em} %
{\renewcommand{\arraystretch}{1.6}%
\FloatBarrier
\begin{longtable}{|c|l|c|c|c|}
 \caption{Spin distribution from individual quarks and total spin
distribution computed in statistical model. We have defined all the
properties in terms of parameters
$\alpha$ and $\beta$.}\\
\hline
 Baryon &Quark spin polarizations and distribution &Calculated Values & Experimental data \cite{19}  \\
\hline
$\Lambda$&$\Delta u=\frac{\alpha}{2}-2\beta$&-0.02&-0.03\\
&$\Delta d=\frac{\alpha}{2}-2\beta$&-0.02&-0.03\\
&$\Delta s=2\alpha+\beta$&0.70&0.74\\
&$I_{1}^{\Lambda}=\frac{1}{4}(\alpha-2\beta)$&0.041&0.027\\\hline

$\Sigma^{+}$&$\Delta u=3\alpha$&0.91&0.98\\
&$\Delta d=0$&$-7.40\times10^{-17}$&-0.02\\
&$\Delta s=-3\beta$&-0.21&-0.29\\
&$I_{1}^{\Sigma^{+}}=\frac{2}{3}\alpha-\frac{1}{6}\beta$&0.191&-\\\hline

$\Sigma^{0}$&$\Delta u=\frac{3}{2}\alpha$&0.46&0.48\\
&$\Delta d=\frac{3}{2}\alpha$&0.46&0.48\\
&$\Delta s=-3\beta$&-0.22&-0.29\\
&$I_{1}^{\Sigma^{0}}=\frac{5}{12}\alpha-\frac{1}{6}\beta$&0.117&-\\\hline

$\Sigma^{-}$&$\Delta u=\frac{1}{6}\alpha-\frac{1}{6}\beta$&0.03&-0.02\\
&$\Delta d=3\alpha$&0.91&0.98\\
&$\Delta s=-3\beta$&-0.22&-0.29\\
&$I_{1}^{\Sigma^{-}}=\frac{1}{6}\alpha-\frac{1}{6}\beta$&0.0389&-\\\hline

$\Xi^{0}$&$\Delta u=-3\beta$&-0.22&-0.29\\
&$\Delta d=0$&0&-0.02\\
&$\Delta s=3\alpha$&0.95&0.98\\
&$I_{1}^{\Xi^{0}}=\frac{1}{2}\alpha-\beta$&0.0838&-\\\hline

$\Xi^{-}$&$\Delta u=-3\beta$&-0.2&-0.020\\
&$\Delta d=\alpha-4\beta$&0.017&-0.29\\
&$\Delta s=3\alpha$&0.95&0.98\\
&$I_{1}^{\Xi^{-}}=\frac{1}{6}\alpha-\frac{1}{6}\beta$&0.0404&-\\\hline
\end{longtable} \FloatBarrier

\section{Discussion of result and conclusion}
Statistical models provides physical simplicity in describing the
various properties of the baryonic states which includes the "sea".
Baryonic structure is considered to be consisting of valence quarks
and sea limited by a few number of quark-antiquark pairs
multiconnected non-perturbatively through gluons. In the present
article, the baryon octet wave function is studied, by looking at
the concept of effective mass and effective charge to analyze
various baryonic properties like magnetic moments, flavor asymmetry
and quark spin distribution.

The explicit numerical values of quark effective mass and quark
effective charge, contributing to the magnetic moments of
$J^{P}=\frac{1}{2}^{+}$ octet baryons are calculated. Effective
quark masses and effective quark charges for quarks u, d, and s are
calculated using fixed inputs for baryon masses (PDG) and
statistical parameters $(\alpha,\beta)$ as input in the respective
formulae. These effective masses and effective charges of u, d and s
are acting as an input to the magnetic moment of
$J^{P}=\frac{1}{2}^{+}$ baryon octet. The contribution of quark
effective charge can be added by taking the concerned parameters as
$\alpha$= 0.248, $\beta$=0.025, $\gamma$= 0.018 to 0.029, as
discussed in the previous section. SU(3) breaking is studied for the
magnetic moment by using a parameter "r" which plays an important
role by providing the basis to understand the extent to which sea
quarks contribute to the structure of the baryon.

We have also investigated the flavor asymmetry of all the octet
baryons. It can be seen from table 1 that, there exist simple
relations between the flavor asymmetries, e.g., the excess of
$\overline{d}$ over $\overline{u}$ in the proton is equal to excess
of $\overline{u}$ over $\overline{d}$ in the neutron, and similarly
for other hyperons. The isospin symmetry leads to these relations
among the flavor asymmetries of octet baryons.

The magnetic moments of the baryon octet is studied in the framework
of the statistical model along with principle of detailed balance in
which the effect of "sea" is taken into account via inclusion of
quark effective mass and quark effective charge. It is interesting
to observe that our results for the magnetic moments of J/P=1/2+
octet particles give a good match with the experimental values
specifically when calculated with quark effective mass (with 2
gluons) are taken whereas magnetic moments deviate when quark
effective charge is considered, as seen in table 2. Though in all
the cases, the contribution of quark sea is quite significant. The
calculated values of magnetic moments with SU(3) breaking is
compared when magnetic moments with SU(3) symmetry is taken and can
be seen in table 3. The listed values shows that the strange quark
contribution to the magnetic moment due to its mass is almost an
order of magnitude smaller than the up and down quarks thus leading
to a very small contribution from the heavy quarks when compared
with the contribution coming from the light quarks. The results with
SU(3) symmetry breaking is not in much agreement with the
experimentally observed values. Plausibly, due to poor role at such
a high energy where s=u=d is applicable.

To appreciate the strange quark in sea, a factor called strangeness
suppression factor is calculated for all the octet particles (in
table 4). This suppression factor suggests the probability of
accommodation of $s\overline{s}$ pairs in sea for singly or doubly
strange octet baryons. Though the data for this factor is
experimentally available only for proton, but we have calculated
this strangeness suppression factor for all the particles in octet.
Hence, this suppression factor suggests that $s\overline{s}$ sea
accomodability enhances for particles with higher strangeness in
their valence part.

Importance of sea with effective mass and charge of the quark
content has been studied and its various effects have been shown
through the above properties, which provide  rich information about
the structure of all the octets thereby motivating experiments for
further inspection. Hence, the validity of the statistical model in
the hadronic structure has been proved for various cases.

\section{Acknowledgement}
The authors gratefully acknowledge the financial support by the
Department of Science and Technology (SB/FTP/PS-037/2014), New
Delhi.








%
%

\end{document}